\documentclass[pra,aps,amsmath,amssymb,amsfonts,reprint,floatfix,twocolumn,nofootinbib,longbibliography,superscriptaddress]{revtex4-2}
\usepackage{amssymb}
\usepackage{bm,mathrsfs}
\usepackage{graphicx}
\usepackage{epsfig}
\usepackage{amsmath,bbm}
\usepackage{amsfonts,amssymb}
\usepackage{times}
\usepackage{verbatim}
\usepackage[sort&compress]{natbib}
\usepackage{amsmath}
\usepackage{bm}
\usepackage{float}
\usepackage{booktabs}  
\usepackage{array}

\allowdisplaybreaks[4]
\usepackage[colorlinks,breaklinks,linkcolor=blue,anchorcolor=blue,citecolor=blue,urlcolor=blue]{hyperref}
\begin{document}

\title{Enhancing ground-state interaction strength of neutral atoms via Floquet stroboscopic dynamics}

\author{Y. Wei}
\affiliation{Center for Quantum Science and School of Physics, Northeast Normal University, Changchun 130024, China}

\author{M. Artoni}
\email{maurizio.artoni@unibs.it }
\affiliation{Department of Engineering and Information Technology, Brescia University, 25133 Brescia, Italy}

\author{G. C. La Rocca}
\email{giuseppe.larocca@sns.it}
\affiliation{European Laboratory for Nonlinear Spectroscopy \& Istituto Nazionale di Ottica del CNR (CNR-INO), 50019 Sesto Fiorentino, Italy}
\affiliation{NEST, Scuola Normale Superiore, 56126 Pisa, Italy}

\author{J. H. Wu}
\email{jhwu@nenu.edu.cn}
\affiliation{Center for Quantum Science and School of Physics, Northeast Normal University, Changchun 130024, China}

\author{X. Q. Shao}
\email{xqshao@nenu.edu.cn}
\affiliation{Center for Quantum Science and School of Physics, Northeast Normal University, Changchun 130024, China}
\affiliation{Institute of Quantum Science and Technology, Yanbian University, Yanji 133002, China}

\begin{abstract}
Neutral atom systems are promising platforms for quantum simulation and computation, owing to their long coherence times. However, their intrinsically weak ground-state interactions pose a major limitation to the advancement of scalable quantum simulation and computation. To address this challenge, we propose an approach to enhancing the ground-state interaction strength of neutral atoms via Floquet modulation of a Rydberg atomic ensemble. Each Floquet period consists of ground-state coupling followed by a pulse driving the transition from the ground state to the Rydberg state. Theoretical analysis and numerical simulations demonstrate that after a defined evolution time, neutral atoms within Rydberg ensembles can collectively form a $W$ state in the ground-state manifold. Even when the Rydberg interaction strength is far below the blockade regime, the fidelity remains remarkably high. Finally, we analyze the application of this scheme in the preparation of single-photon sources. In general, our proposed mechanism offers an efficient and highly controllable method for quantum state preparation within the Rydberg atomic ensembles, significantly enhancing the accuracy and stability of quantum state engineering while providing a well-controlled quantum environment for single-photon generation.
\end{abstract}

\maketitle

\section{Introduction}\label{sec1}
Ground-state neutral atoms exhibit long coherence times and excellent isolation from environmental noise, making them promising candidates for scalable quantum technologies~\cite{PhysRevLett.104.010502,kaufman2015entangling,Henriet2020quantumcomputing,PhysRevA.105.032417,Shi_2022,Baroni2024,zqkr-8991}. Meanwhile, coupling ground states to Rydberg levels offers a complementary advantage: strong, long-range interactions via van der Waals (vdW) or dipole-dipole interactions, which enable fast entangling operations through the Rydberg blockade effect~\cite{urban2009observation,RevModPhys.82.2313,PhysRevLett.112.073901,PhysRevA.101.042328,PhysRevLett.128.033201,PhysRevLett.134.053604}. The Rydberg blockade provides a powerful mechanism for engineering strong, controllable interactions and creating collective quantum states in which a single excitation is coherently shared throughout the atomic ensemble~\cite{Kumlin_2023,10.1063/5.0211071,PhysRevLett.133.213601}. Such collective control has opened broad application prospects in quantum optics~\cite{peyronel2012quantum,PhysRevA.89.063407,chang2014quantum,Bai:16,busche2017contactless,PhysRevX.12.021034,photonics9040242}, quantum computing~\cite{Saffman_2016,10.1116/5.0036562,Wu_2021}, quantum communication~\cite{PhysRevA.95.022317,10.1063/1.5031033,PhysRevLett.123.140504,yang2022sequential,PhysRevLett.128.060502}, and single-photon sources~\cite{PhysRevA.66.065403,PhysRevLett.107.093601,Distante2017,PhysRevLett.121.123605,Bastidas_2018,PhysRevLett.125.073602,Yang:22,shi2022high,PhysRevA.109.013705,PhysRevA.109.013710,cmpn-jfqr}. With the development of experimental technology, the coherent manipulation of Rydberg atomic ensembles is driving a deeper understanding of collective quantum dynamics and enabling new opportunities for scalable and high-performance quantum technologies~\cite{Mukherjee_2011,dudin2012observation,Letscher_2017,PhysRevA.98.052324,PhysRevLett.125.263605,PhysRevA.101.053432,PhysRevLett.126.233404,Ding2022,PhysRevLett.131.123201,w3x9-ll79}.

The finite lifetime and strong decoherence effects inherent to Rydberg states present significant challenges for high-fidelity quantum operations. This naturally raises the question: Can sufficiently strong and controllable interactions be engineered while atoms remain in their ground state? In this context, finding ways to achieve effective interactions without exciting high-energy states is especially important. Just as the Rydberg blockade allows only a single collective excitation in Rydberg states, strong interactions between ground states can also give rise to a ground-state blockade mechanism~\cite{PhysRevLett.96.063001,weimer2010rydberg,Jau2016,PhysRevA.96.012328,PhysRevLett.125.073601,Xu_2022}. Consequently, the system is deterministically prepared in the target symmetric collective state in the ground-state manifold. This symmetric collective state is of great significance for photon storage~\cite{corzo2019waveguide,PhysRevLett.129.253601}, single-photon sources~\cite{grafe2014chip,PhysRevLett.117.223001,Khazali_2017,PhysRevA.105.062408,PhysRevResearch.7.013238}, quantum state transfer~\cite{PhysRevA.103.052201,s7j9-z7f8}, and other quantum information technologies~\cite{Cotrufo:19,c47j-cw46}. Compared to the complexity and strong interactions inherent in Rydberg states, the symmetric collective state of ground states provides a more precise and stable quantum platform, making it particularly promising for applications in quantum computing and quantum simulation~\cite{PhysRevLett.92.213601,PhysRevLett.118.253601,PhysRevResearch.3.033287,evered2023high}.

In this work, we utilize Floquet stroboscopic dynamics to prepare a $W$ state within the ground-state manifold of a Rydberg atomic ensemble~\cite{morton2006bang,PhysRevA.75.063424,SHAO20111099,PhysRevA.85.011605,Bukov04032015,PhysRevLett.117.250401,PhysRevA.95.062339,PhysRevA.109.062414,5qhh-322q}. The system evolves from the ground state $|G\rangle=|g_1 g_2...g_\mathbb{N} \rangle$ to the symmetric collective state $|W_\mathbb{N}\rangle= {1/\sqrt{\mathbb{N}} \textstyle \sum_{i=1}^{\mathbb{N}}} |g_1 g_2...e_i...g_\mathbb{N} \rangle$, while the double excitations remain unpopulated at stroboscopic time points. This indirectly induces an effective interaction between ground-state neutral atoms, leading to a strong blockade effect that enables high-fidelity preparation of the $W$ state. Finally, we analyze the application of the scheme to single-photon generation. Our approach remains effective not only within the Rydberg blockade radius but also well beyond it, offering clear advantages over conventional blockade-based protocols.

The remainder of this paper is structured as follows. In Sec.~\ref{sec2}, we introduce the theoretical model and describe the physical mechanism of the proposed scheme. In Sec.~\ref{sec3}, we present a comprehensive performance assessment, including the dependence on the Rydberg interaction strength and the robustness of the protocol against various experimentally relevant imperfections, such as Rydberg-state decay (including blackbody radiation), laser phase and amplitude noise, detuning errors, and Doppler-induced collective dephasing. We also compare our scheme with several representative approaches. In Sec.~\ref{sec4}, we analyze the application of our scheme to the realization of single-photon sources. In Sec.~\ref{sec5}, we discuss its implementation in a Rydberg superatom. In Sec.~\ref{sec6}, we provide the corresponding energy level diagram and transitions. Finally, We conclude our work in Sec.~\ref{sec7}.

\section{Theoretical Model and Physical Mechanism}\label{sec2}
\subsection{System Hamiltonian and Theoretical Framework}
\begin{figure}
	\centering
	\includegraphics[width=1\linewidth]{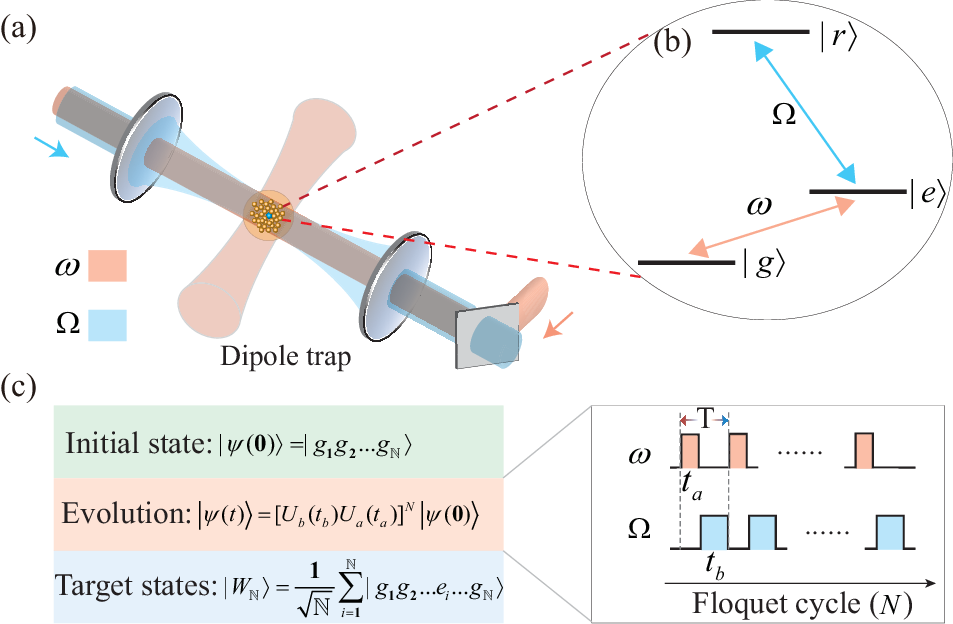}
	\caption{(a) An ensemble of cold $^{87}\text{Rb}$ atoms is confined in a dipole trap. (b) Schematic diagram of single atomic energy level configuration with ground state $|g\rangle \equiv |5S_{1/2},F=1,m_F=0\rangle $ and $|e\rangle \equiv |5S_{1/2},F=2,m_F=0\rangle $, the Rydberg state is $|nS_{1/2},m_j=1/2\rangle$. (c) The evolution and corresponding pulse sequence of the system under periodic driving.}\label{fig1}
\end{figure}

As illustrated in Fig.~\hyperref[fig1]{1(a)}, we consider a cold ensemble of $^{87}\text{Rb}$ atoms confined in an optical dipole trap. The collective blockade radius is given by $R_b=[|C_6|/(\hbar \sqrt{\mathbb{N}}\Omega)]^{1/6}$. Here, $C_6$ is the vdW dispersion coefficient, and $\Omega$ denotes the effective Rabi frequency between the ground state and the Rydberg state. For simplicity, we adopt $\hbar=1$ in the subsequent analysis. The relevant atomic energy levels are shown in Fig.~\hyperref[fig1]{1(b)}. In the interaction picture,the ground states $|g\rangle$ and $|e\rangle$ are coupled by a coherent driving field, which can be implemented either as a direct microwave transition or as an effective two-photon Raman coupling, with Rabi frequency $\omega$, described by the Hamiltonian
\begin{align}\label{eq1}
H_a &= \sum_{i=1}^{\mathbb{N}} \frac{\omega}{2} |e_i\rangle\langle g_i| + \mathrm{H.c.}
\end{align}
Meanwhile, the ground state $|e\rangle $ can be further excited to the Rydberg state $|r\rangle$ by another driving field with two-photon Rabi frequency $\Omega$, leading to the Hamiltonian
\begin{align}\label{eq2}
H_b &= \sum_{i=1}^{\mathbb{N}} \frac{\Omega}{2} |r_i\rangle\langle e_i| + \mathrm{H.c.}+\frac{1}{2}\sum_{i\ne j} U_{rr} |r_i r_j\rangle\langle r_i r_j|.
\end{align}
Here, $U_{rr}$ denotes the Rydberg-mediated interaction that occurs when two atoms are simultaneously excited to the Rydberg state. This interaction is of the vdW type and can be expressed as $U_{rr} = -C_6/d_{i,j}^6$, where $d_{i,j}$ is the distance between the $i$-th and $j$-th atoms. The operation is performed periodically over a total of $N$ cycles, with each cycle having a period of $T = t_a + t_b$. Each cycle consists of two successive stages: during the interval $0 \leq t < t_a$, the dynamics are governed by the Hamiltonian $H_a$, while in the subsequent interval $t_a \leq t < t_b$, the system evolves under the Hamiltonian $H_b$. For convenience, the durations are chosen as $t_a=\pi/({\sqrt{\mathbb{N}} N \omega})$ and $t_b=4\pi/\Omega$, respectively, corresponding to a collective $\pi$ pulse on the $|g\rangle \leftrightarrow |e\rangle$ transition across all atoms and a $4\pi$ pulse on the $|e\rangle \leftrightarrow |r\rangle$ transition~\cite{Bukov04032015,5qhh-322q}. The stroboscopic evolution of the system after each period is described by the Floquet evolution operator $U_N (T)$, which governs the dynamics of discrete time. The explicit expression for $U_N (T)$ is given as follows:
\begin{align}\label{eq3}
U_N (T)=[ U_{b}(t_b) U_{a} (t_a)]^N,
\end{align}
$U_a(t_a)$ and $U_b(t_b)$ denote the evolution operators corresponding to the Hamiltonians $H_a$ and $H_b$, respectively. This equation describes the full dynamical evolution of the system; however, its analytical solution is generally intractable. In the large-$N$ limit, the dynamics can be characterized by unitary kicks~\cite{PhysRevLett.89.080401,morton2006bang,SHAO20111099}, the effective evolution in the large-$N$ limit can be written as 
\begin{align}\label{eq4}
\lim_{N \to \infty} U_N (T) &\equiv \mathcal{U}_N (T)\notag \\
& \sim U_{b} (t_b )^N e^{-i H_z N t_a}\notag  \\
&=\exp \left[-i \sum_n (\lambda_n P_n \frac{t_b}{T} + P_n H_a P_n \frac{t_a}{T}) N T\right].
\end{align}
$H_z$ is referred to as the Zeno Hamiltonian. $P_n$ represents the spectral projection of $U_b$ corresponding to the eigenvalue $\lambda_n$. For which an analytical expression is derived in Appendix~\ref{Appendix_A}.

\begin{figure*}
	\centering
	\includegraphics[width=0.8\linewidth]{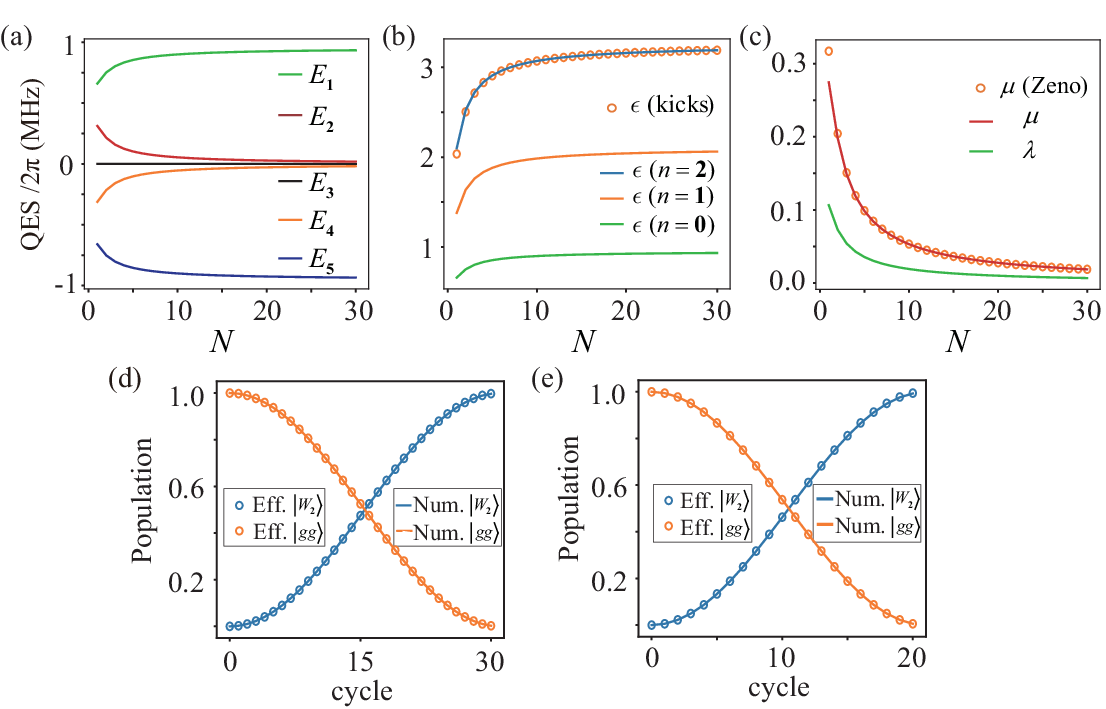}
	\caption{(a) Floquet quasienergy spectrum of the system. (b-c) Two representative regions extracted from panel (a), showing the comparison between the Floquet quasienergy spectrum and the effective coupling obtained from the unitary kicks Hamiltonian for different driving periods $N$, in accordance with the quasienergy periodicity relation $\epsilon \equiv \epsilon + 2n\pi/T$. (d) and (e) show population of the states for a two-atom system under $N=30$ and $N=20$ cycles, respectively. We set $\omega/2\pi=1$~MHz, $\Omega= 5 \omega$, and $U_{rr}= 45 \Omega$.}\label{fig2}
\end{figure*}

\subsection{Mechanism of Enhanced Ground-State Interaction and Ground-State Blockade}

To illustrate the mechanism, we start with a simple two-atom system and then generalize it to larger ensembles. Considering the existence of the Rydberg blockade, we can write five basis vectors in the symmetric state: \{$|gg\rangle$, $|W_2\rangle=(|ge\rangle+|eg\rangle)/\sqrt{2}$, $|ee\rangle$, $|T_2\rangle=(|gr\rangle+|rg\rangle)/\sqrt{2}$, $|D_2\rangle=(|re\rangle+|er\rangle)/\sqrt{2}$\}. Therefore, the corresponding coupled Hamiltonian can be written as
\begin{align}\label{eq5}
H^{(1)}_{a}&=\frac{\sqrt{2}\omega}{2}  |W_2\rangle \langle gg| +\frac{\sqrt{2}\omega}{2}  |ee\rangle \langle W_2| + \frac{\omega}{2} |D_2\rangle \langle T_2|+ \mathrm{H.c.},\notag\\
H^{(1)}_{b}&=\frac{\sqrt{2}\Omega}{2}  |ee\rangle \langle D_2| + \frac{\Omega}{2} |T_2\rangle \langle W_2|+ \mathrm{H.c.}
\end{align}
From the perspective of unitary kicks, because the system is initialized in $|gg\rangle$, the stroboscopic evolution is confined to the $H_z$ subspace, and the transition to the doubly excited state $|ee\rangle$ is strongly suppressed. The effective Hamiltonian can therefore be written as $H^{\text{eff}}_Z=H_z t_a /(t_a+t_b)$, where $H_z =\frac{\sqrt{2} \omega}{2} (|gg\rangle \langle W_2|+ \mathrm{H.c.})$. In other words, the unitary kicks effectively decouple $|ee\rangle$ from the dynamics. In fact, from a numerical perspective, the above dynamical process is independent of the number of driving periods $N$. For any value of $N$, the numerical results can be well explained by the effective Hamiltonian derived from the Floquet theorem. The corresponding effective Hamiltonian can be written as $H_\mathrm{eff}=i \mathrm{ln} (e^{-iH^{(1)}_{b}t_b} e^{-iH^{(1)}_{a}t_a})/T$, which allows us to further verify the precision of the analytical results through numerical simulations (see Appendix~\ref{Appendix_B}). To verify the validity of this approach, we further consider finite values of $N$ and compare the resulting quasienergy spectrum with the Floquet theorem, as shown in Fig.~\hyperref[fig2]{2(a)}. Each curve represents the quasienergy corresponding to an eigenstate, which is expressed as a linear superposition of the same set of basis vectors with different weights for different values of $N$. It can be seen that, for small $N$, the five branches of the quasienergy spectrum are close to each other. With increasing $N$, the structure of the energy gap gradually opens up, with $E_1$ and $E_5$ shifting towards higher and lower energies, respectively, while $E_2$, $E_3$, and $E_4$ converge to zero. In the large-$N$ limit, the eigenstate structure exhibits a clear pattern: the eigenstates associated with $E_1$ and $E_5$ are primarily composed of $|ee\rangle$ and $|D_2\rangle$, and can be approximately written as $|\Psi_{1,5}\rangle \approx c_1 |D_2\rangle \mp c_2 |ee\rangle$; similarly, the eigenstates corresponding to $E_2$ and $E_4$ are dominated by $|gg\rangle$ and $|W_2\rangle$, expressed as $|\Psi_{2,4}\rangle \approx c_3 |gg\rangle \mp c_4 |W_2\rangle$, while the eigenstate associated with $E_3$ is essentially given by $|T_2\rangle$ and $|\Psi_3\rangle \approx |T_2\rangle$, with all other components suppressed below $10^{-2}$, rendering their influence on the dynamics negligible. As $N$ increases, the effective dimensionality of the eigenstates decreases significantly, and the spectral structure becomes more distinguishable. Reconstructing $H_{\mathrm{eff}}=\sum_i E_i|\Psi_i\rangle\langle\Psi_i|$ and projecting it onto the reduced subspace spanned by $\{|gg\rangle, |W_2\rangle, |ee\rangle, |D_2\rangle\}$, we identify the dominant effective couplings that govern the dynamics. In this approximation, the quasienergy branch $E_1$ is equivalent to the coupling between $|ee\rangle$ and $|D_2\rangle$ with strength $\epsilon$, and $E_2$ is equivalent to the coupling between $|gg\rangle$ and $|W_2\rangle$ with strength $\mu$. These effective couplings capture the essential structure of the quasienergy spectrum and provide a clear physical interpretation of the underlying dynamics.

Furthermore, Figs.~\hyperref[fig2]{2(b)} and \hyperref[fig2]{2(c)} present the analytical results of the unitary kicks and the numerical results corresponding to the quasienergy spectrum shown in Fig.~\hyperref[fig2]{2(a)}. The green line in Fig.~\hyperref[fig2]{2(b)} represents the effective coupling strength $\epsilon$ between $|ee\rangle$ and $|D_2\rangle$, corresponding to the quasienergy branch $E_1$ shown in Fig.~\hyperref[fig2]{2(a)}. The orange and blue curves in Fig.~\hyperref[fig2]{2(b)} represent the quasienergy periodicity relation $\epsilon \equiv \epsilon + 2n\pi/T$ for different $n$, {where $n$ is an integer indexing the periodic Floquet branches~\cite{PhysRevLett.134.063602,8hkk-2c6d},} while the orange line with circular markers represents the analytical results obtained from the unitary kicks. Owing to the intrinsic periodicity of the Floquet quasienergies, the analytical results coincide with the $n=2$ branch of the Floquet spectrum under the $4\pi$-pulse condition. As $N$ increases, the $|ee\rangle \to |D_2\rangle$ transition becomes significantly stronger, analogous to the enhancement of dipole-dipole interactions in Rydberg atomic ensembles. The solid red line in Fig.~\hyperref[fig2]{2(c)} represents the effective coupling strength $\mu$ between $|gg\rangle$ and $|W_2\rangle$, corresponding to the quasienergy branch $E_2$ shown in Fig.~\hyperref[fig2]{2(a)}, while the orange line with circular markers represents the analytical results obtained from the unitary kicks. The green line denotes the numerical result of the Floquet theorem for the effective coupling strength $\lambda$ associated with the transition between $|ee\rangle$ and $|W_2\rangle$. We can observe that as $N$ increases, $\lambda$ and $\mu$ gradually decrease. This indicates that when the time scale $t_a$ is small, there exists a strong coupling interaction between the $|ee\rangle$ and $|D_2\rangle$ states, while the coupling from the $|ee\rangle$ to the $|W_2\rangle$ state is relatively weak. This result is similar to Autler-Townes splitting, which prevents the transition from $|gg\rangle$ to $|ee\rangle$, thus evolving only from $|gg\rangle$ to $|W_2\rangle$. Usually, ground-state interactions are relatively weak at micrometer-scale interatomic distances. In other words, our scheme indirectly enhances the interaction strength between ground states. {This induced strong coupling gives rise to a ground-state blockade, in which the simultaneous population of multiple atoms in the $|e\rangle$ state is effectively suppressed, thereby enabling the preparation of entangled states within the ground-state manifold.}

Having analyzed the static quasienergy spectrum, we now demonstrate the blockade effect through dynamical simulations. As shown in Fig.~\hyperref[fig2]{2(d)}, we numerically simulate a two-atom system governed by Eqs.~(\ref{eq3}) and (\ref{eq4}) to assess the feasibility of realizing ground-state blockade. The blue and yellow curves denote the populations of the states $|W_2\rangle$ and $|gg\rangle$, respectively. Solid lines correspond to the full numerical evolution under Eq.~(\ref{eq3}), while curves with circular markers represent the effective dynamics described by Eq.~(\ref{eq4}).
We first consider the dynamical evolution over $N=30$ driving cycles. In this case, the population of the target state $|W_2\rangle$ reaches approximately $99.6\%$, demonstrating excellent agreement between the full and effective descriptions and confirming the validity of the effective evolution for sufficiently large $N$. Figure~\hyperref[fig2]{2(e)} presents the results for a reduced number of cycles, $N=20$, where the scheme remains robust and achieves nearly complete population transfer from $|gg\rangle$ to $|W_2\rangle$. Both the numerical and effective solutions exhibit consistent behavior, yielding a final $|W_2\rangle$ population of about $99.2\%$.
These results validate our theoretical analysis and confirm that ground-state blockade can be efficiently realized via enhanced ground-state interactions within the proposed scheme. To ensure numerical accuracy, all subsequent simulations are based on the full numerical solution of Eq.~(\ref{eq3}). In the main text, we present only the stroboscopic dynamics, while Appendix~\ref{Appendix_C} provides the corresponding continuous-time evolution. All numerical calculations are performed using the QuTiP package in Python~\cite{JOHANSSON20121760,JOHANSSON20131234}.

\section{Performance Assessment}\label{sec3}
\subsection{Fidelity and Quasienergy Spectrum Analysis}
\begin{figure}
	\centering
	\includegraphics[width=1\linewidth]{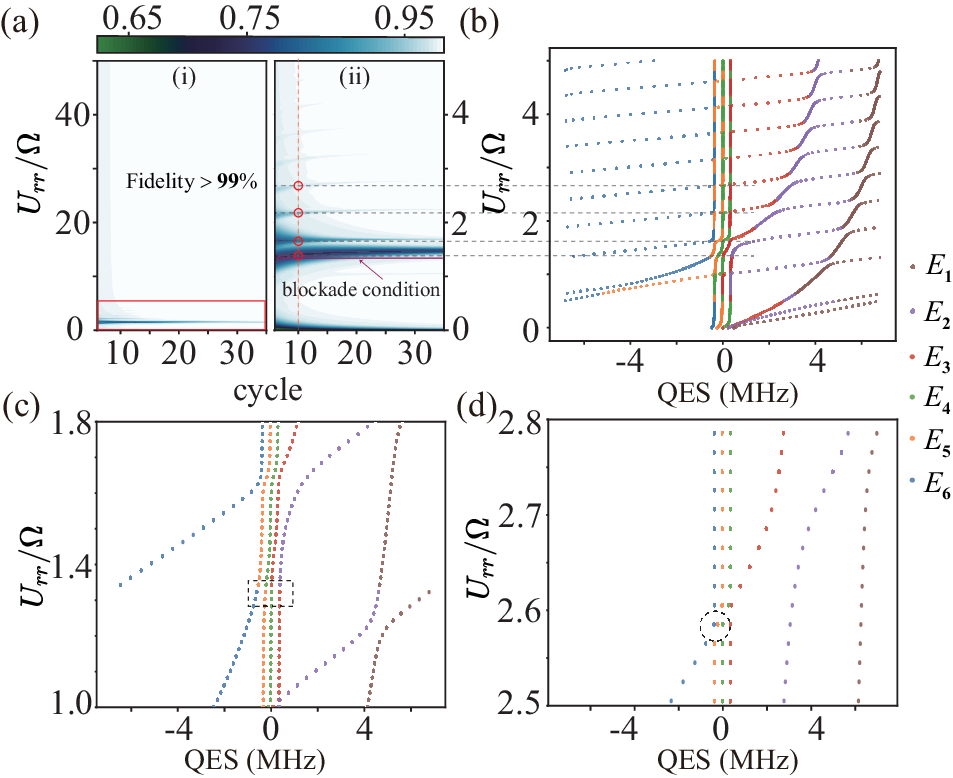}
	\caption{(a) Fidelity of the state $|W_2\rangle$ versus $U_{rr}/\Omega$ and cycle number $N$, showing high overall robustness ($F>99\%$) but revealing several sharp, $N$-independent resonant leakage channels, as shown in the broad scan (i) and detailed view (ii). (b) Quasienergy spectrum as a function of $U_{rr}/\Omega$, revealing a distinct avoided crossing. (c-d) Detailed views of the quasienergy spectrum as a function of $U_{\mathrm{rr}}/\Omega$. Other parameters are the same with Fig.~\ref{fig2}.}\label{fig3}
\end{figure}

So far, the analysis has been based on idealized theoretical assumptions. To explore the feasibility of our scheme in broader scenarios and to examine its dependence on the Rydberg blockade effect, we analyze its performance under different Rydberg interaction strengths with two atoms. Figure~\hyperref[fig3]{3(a)} shows the fidelity distribution of the preparation of the state $|W_2\rangle$, with the vertical and horizontal axes representing the total number of pulse periods $N$ and the Rydberg interaction strength $U_{rr}$. Consequently, each vertical slice in the figure corresponds to a complete dynamical simulation with a different total evolution time $N T$. The figure reveals two key features, visible in both the full parameter scan (panel i) and the locally magnified region (panel ii).
Firstly, in the vast majority of the parameter space, the scheme achieves fidelities that exceed $99\%$ (white regions). More interestingly, in the weak Rydberg interaction regime, the scheme still achieves efficient state preparation, demonstrating its robustness to reduced interaction strength.
Secondly, a low-fidelity vertical canyon emerges in certain parameter regions, most prominently around $U_{rr}/\Omega \approx 1.25$, as shown in panel (ii). This systematic feature persists in both small and large $N$, indicating a strong dip in fidelity at this interaction strength. In addition, we conducted analyzes with three and four atoms in Appendix~\ref{Appendix_D}, and the results were consistent with those obtained for two atoms.

To understand the fidelity decrease in these regions, as shown in Fig.~\hyperref[fig3]{3(b)}, we analyzed the quasienergy spectrum of the system. At the point where fidelity decreases most significantly, around $U_{rr}/\Omega \approx 1.25$, the quasienergies $E_3$, $E_5$, and $E_6$ are closely spaced, leading to pronounced nonadiabatic transitions during evolution, as shown in Fig.~\hyperref[fig3]{3(b)} and its zoomed-in view in Fig.~\hyperref[fig3]{3(c)}. The corresponding eigenstates are $|\psi_3\rangle \approx 0.65|gg\rangle + 0.72|W_2\rangle$, $
|\psi_5\rangle \approx 0.71|gg\rangle - 0.57|W_2\rangle$, and $|\psi_6\rangle \approx 0.16 e^{-0.14i}|gg\rangle + 0.37e^{3.00i}|W_2\rangle + 0.69|T_2\rangle + 0.46e^{0.02i}|D_2\rangle + 0.33e^{-2.41i}|rr\rangle$. Ideally, the system should adiabatically follow the $|gg\rangle \rightarrow |W_2\rangle$ pathway. However, small energy gaps induce transitions from $|\psi_3\rangle$ and $|\psi_5\rangle$ to $|\psi_6\rangle$, which contain significant components $|ee\rangle$, $|D_2\rangle$, and $|rr\rangle$. This transition results in a noticeable enhancement of the $|ee\rangle$ population. For other regions where the fidelity does not exhibit a significant decrease, such as at $U_{rr}/\Omega \approx 2.58$, an avoided crossing occurs between $E_5$ and $E_6$, as shown in Fig.~\hyperref[fig3]{3(b)} and its zoomed-in view in Fig.~\hyperref[fig3]{3(d)}. The corresponding eigenstates are $|\psi_5\rangle \approx -0.63|gg\rangle + 0.68|W_2\rangle + 0.29e^{-1.62i}|rr\rangle$, $
|\psi_6\rangle \approx 0.29e^{1.66i}|gg\rangle + 0.18e^{-1.48i}|W_2\rangle + 0.18e^{-1.36i}|ee\rangle+ 0.26e^{1.65i}|T_2\rangle + 0.22e^{-1.50i}|D_2\rangle + 0.85|rr\rangle$. The reduced energy gap leads to strong mixing between the two eigenstates, significantly increasing the nonadiabatic transition probability in the avoided-crossing region. As a result, part of the population is transferred from $|\psi_5\rangle$ to $|\psi_6\rangle$, which has higher weights $|ee\rangle$ and $|rr\rangle$, thus further enhancing the population of $|ee\rangle$. The independence from the number of driving cycles $N$ confirms that this effect originates from an intrinsic energy matching condition of the Floquet Hamiltonian, rather than from a dynamical process dependent on timescale. Apart from the dark blue region, fidelity remains remarkably close to unity. Interestingly, even for values of the Rydberg interactions much smaller than those corresponding to the blockade regime, the fidelity remains very large. Notice that for very small interaction strengths, we may maintain large fidelities by increasing the number of cycles. In contrast, most previous schemes strongly rely on the Rydberg blockade effect~\cite{Yang:22,PhysRevA.109.013705,PhysRevLett.125.073602}, whereas our method does not. Although the performance of the scheme is reduced when $U_{rr}/\Omega \approx 1.25$, its key advantage lies in the fact that, under periodic driving, fluctuations in atomic positions cause the Rydberg interaction strength to vary rather than remain fixed at a specific value during each dynamical evolution. The combined effect of periodic driving and positional fluctuations effectively mitigates the detrimental influence of unfavorable Rydberg interaction strengths. This feature makes our scheme particularly suitable for future implementations in atomic arrays with well-defined interatomic spacing~\cite{Labuhn2016,doi:10.1126/science.aah3778,Bekenstein2020,Ebadi2021,Scholl2021,Graham2022,Srakaeew2023,PhysRevLett.131.170601,PRXQuantum.4.010316,Chen2023,Bluvstein2024,PhysRevLett.133.233003,Zhang2025,Manetsch2025,2ym8-vs82,r54t-myhc}.

\begin{figure}
	\centering
	\includegraphics[width=1\linewidth]{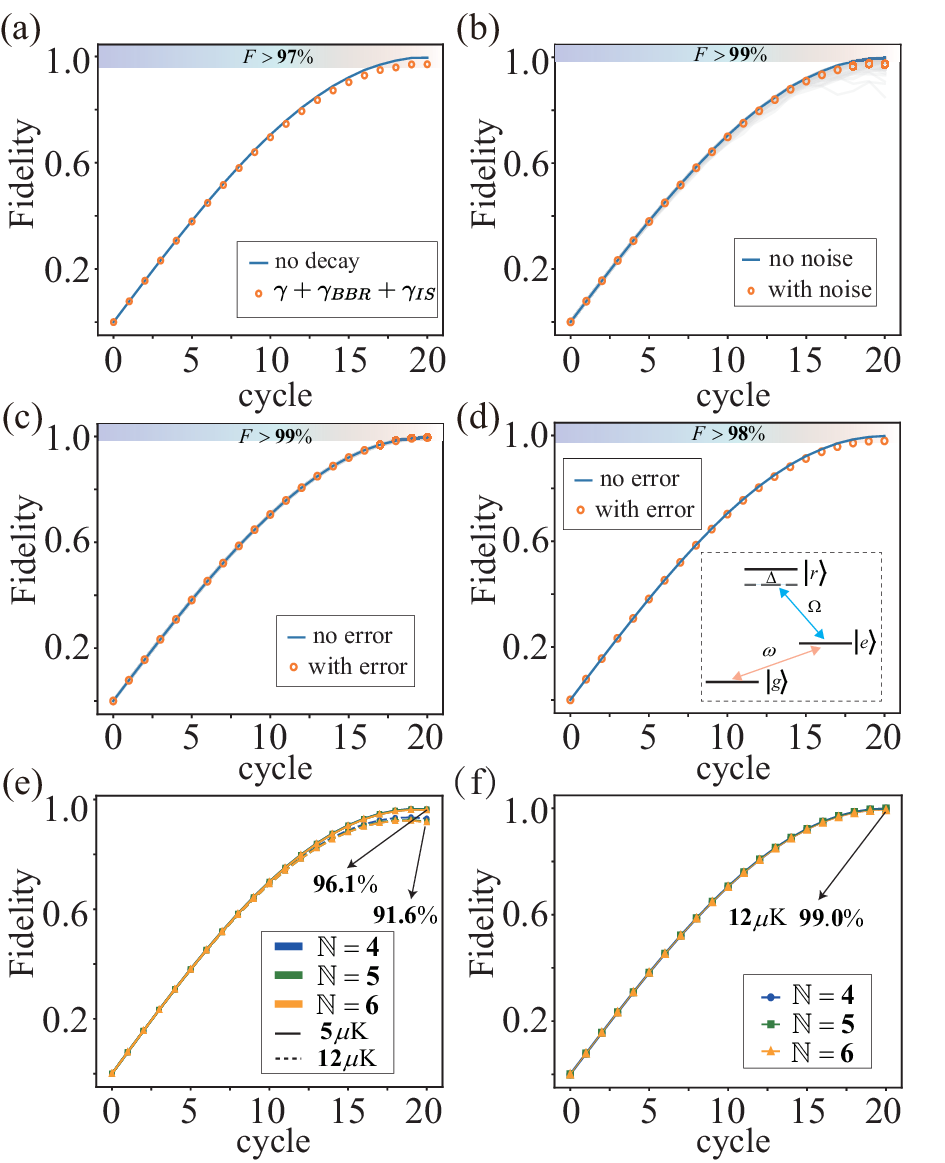}
	\caption{ { Fidelity of the target state $|W_{\mathbb{N}}\rangle$ as a function of the cycle number under different noise and decoherence conditions. (a) Fidelity of the target state $|W_{\mathbb{N}}\rangle$ for $\mathbb{N}=4$ atoms in the presence of atomic position disorder, and the influence of additional decoherence effects including spontaneous decay, blackbody-radiation-induced decay, and intermediate-state scattering. (b--d) Fidelity for $\mathbb{N}=4$ atoms in the presence of different technical noises: (b) Rabi amplitude noise, (c) laser phase noise, and (d) laser detuning noise. (e) Fidelity for different atom numbers $\mathbb{N}=4,5,6$ in the presence of Doppler-induced dephasing at atomic temperatures of $5\,\mu\mathrm{K}$ and $12\,\mu\mathrm{K}$. (f) Same as panel (e), but with an alternating two-photon driving scheme in successive cycles to suppress the accumulated Doppler-induced dephasing. Other parameters are the same with Fig.~\ref{fig2}.}}\label{fig4}
\end{figure}

\subsection{Robustness Against Experimental Noise}

Building on the mechanism described above, we now extend the analysis from a two-atom system to a four-atom system and validate the scheme under realistic conditions by examining the impact of key experimental imperfections, {including static positional disorder of atoms during the driving cycles, spontaneous emission from the Rydberg state $|r\rangle$, blackbody-radiation-induced decay, scattering via the intermediate state, laser phase error, Rabi amplitude noise, laser detuning error, as well as Doppler-induced frequency shifts.} We point out that although our numerical calculations are restricted to the case of $\mathbb{N}=4$ due to the large Hilbert space, the proposed scheme is equally applicable to larger systems ($\mathbb{N}>4$).

{
\textbf{Intrinsic atomic decay and disorder.---}
To evaluate the robustness of the protocol under realistic experimental conditions, we incorporate intrinsic decoherence and static disorder into the system dynamics. The dominant noise sources include spontaneous decay of the Rydberg state, blackbody-radiation-induced transitions among Rydberg levels, and effective dephasing arising from intermediate-state scattering during the two-photon excitation process. The resulting dynamics are described by the Lindblad master equation
\begin{align}
\dot{\rho}
=&\sum_{i=1}^{\mathbb{N}}
\Bigg[
\frac{\gamma}{2}\mathcal{D}\!\left[|e_i\rangle\langle r_i|\right]\rho
+\frac{\gamma}{2}\mathcal{D}\!\left[|g_i\rangle\langle r_i|\right]\rho
\notag\\
&\quad
+\gamma_{\mathrm{BBR}}
\mathcal{D}\!\left[|p_i\rangle\langle r_i|\right]\rho
+\gamma_{\mathrm{IS}}
\mathcal{D}\!\left[|r_i\rangle\langle r_i|\right]\rho
\Bigg],
\end{align}
where $\mathcal{D}[o]\rho = o\rho o^{\dagger} - (o^{\dagger}o\rho + \rho o^{\dagger}o)/2$ is the standard Lindblad superoperator. Here, the Rydberg state is $|r\rangle = |80S_{1/2}\rangle$, with spontaneous decay $\gamma/2\pi = 0.28~\mathrm{kHz}$, blackbody-induced transitions $\gamma_{\mathrm{BBR}}/2\pi = 0.50~\mathrm{kHz}$, and effective dephasing $\gamma_{\mathrm{IS}}$ from intermediate-state scattering. For the two-photon excitation scheme, the intermediate-state scattering rate is given by~\cite{evered2023high}
\begin{equation}
\gamma_{\mathrm{IS}} = \frac{\Omega_{780}^2 + \Omega_{480}^2}{4\Delta_p^2} \Gamma_e ,
\end{equation}
where $\Gamma_e = 2\pi \times 6~\mathrm{MHz}$ is the natural linewidth of the intermediate state. The single-photon Rabi frequencies are $\Omega_{780} = 2\pi \times 237~\mathrm{MHz}$ and $\Omega_{480} = 2\pi \times 303~\mathrm{MHz}$, and the detuning is $\Delta_p = 2\pi \times 7.8~\mathrm{GHz}$. 
Based on the above scheme, we first examine the impact of interaction inhomogeneity, which is particularly relevant for atomic ensembles where the van der Waals interaction strength $U_{rr}$ fluctuates due to thermal motion or shot-to-shot positional variations. In our simulations, $U_{rr}$ is randomly sampled in each driving cycle from a uniform distribution over $[0,\,2\pi \times 200~\mathrm{MHz}]$. The resulting fidelity,
$F(t)=\langle W_\mathbb{N}|\rho(t)|W_\mathbb{N}\rangle$,
is shown in Fig.~\hyperref[fig4]{4(a)}. Remarkably, despite this strong cycle-to-cycle disorder, the final fidelity of the target state $|W_\mathbb{N}\rangle$ consistently exceeds $99.1\%$.
We next perform a more comprehensive simulation that simultaneously incorporates all dominant decoherence mechanisms at room temperature ($300~\mathrm{K}$), including spontaneous decay of the Rydberg state, blackbody-radiation-induced transitions, and scattering via the intermediate state involved in the off-resonant two-photon excitation. The intermediate-state scattering is modeled using the standard treatment of off-resonant Raman processes~\cite{RevModPhys.82.2313,PhysRevX.5.031015,evered2023high}. Even under the combined influence of these dissipation channels, the fidelity of the target state remains as high as $97.5\%$.
This robustness can be attributed to the cycle-to-cycle fluctuations of $U_{rr}$, which prevent the system from residing near the resonant condition $U_{rr}/\Omega \approx 1.25$ for extended durations, thereby strongly suppressing leakage through resonant channels. These results demonstrate that the protocol is resilient against both intrinsic decoherence and realistic dynamical disorder encountered in experiments.
}

{\textbf{Laser Rabi amplitude noise and phase error.---} In realistic experiments, imperfections in the driving fields play an important role. Laser instability and technical noise induce stochastic fluctuations in the driving-field intensity, which can be effectively modeled as Rabi-amplitude noise~\cite{PRXQuantum.4.020336,PRXQuantum.6.010331,PhysRevA.111.022420}.Accordingly, the Rabi frequency acting on atom $i$ is taken as $\Omega_i = [1+\epsilon]\Omega$. Within a realistic error budget, the Rabi-amplitude noise $\epsilon$ is modeled as a random variable following the Gaussian distribution $\mathcal{N}(0, \sigma_\epsilon^2)$ with a standard deviation of $\sigma_\epsilon = 0.01$ (corresponding to $1\%$ amplitude noise), representing a noise level that is well within the capabilities of current experimental amplitude-control techniques. By averaging over 50 independent noise realizations, we obtain the fidelity $F(t)$ shown in Fig.~\hyperref[fig4]{4(b)}. The results show that the protocol is only weakly affected by Rabi-amplitude noise, with the final fidelity remaining above $97.1\%$.
We further consider laser phase noise, which arises from both quantum noise and technical fluctuations and leads to deviations from an ideal coherent drive. The laser phase can be written as $\varphi(t)=\varphi_0+\delta\varphi(t)$, resulting in a modified Rabi frequency $\Omega \rightarrow \Omega e^{i\delta\varphi(t)}$. Within the error margins reported in current experiments~\cite{PhysRevLett.121.123603,PhysRevLett.124.033603,PhysRevA.105.042430}, a typical phase fluctuation of $\delta\varphi = 0.1\pi$ is assumed. In the simulations, this effect is modeled as a random dynamical phase error, and the results are obtained by averaging over 50 independent noise realizations. The corresponding fidelity is shown in Fig.~\hyperref[fig4]{4(c)}. It is found that such phase noise has a negligible impact on the protocol performance.
}

{\textbf{Laser detuning errors.---} Laser detuning errors constitute a major source of systematic imperfections in quantum optical and quantum computing platforms. Such detuning shifts the effective transition frequency and can weaken the Rydberg blockade.} In Fig.~\hyperref[fig4]{4(d)}, The evolution of the fidelity is simulated in the presence of detuning errors randomly sampled from the range $[-2\pi \times 30~\mathrm{kHz},\, 2\pi \times 30~\mathrm{kHz}]$, with the results averaged over 50 independent realizations. It is found that the fidelity remains as high as $98.8\%$ in this range of detuning errors. This indicates that the protocol is robust against detuning imperfections, provided that the errors remain within experimentally relevant bounds.

{\textbf{Doppler dephasing.---} Doppler shifts associated with the coupling between the ground states $|g\rangle$ and $|e\rangle$ are negligible, as the coupling is mediated either by microwave fields with long wavelengths or by co-propagating optical fields with nearly identical wave vectors, resulting in a strongly suppressed effective wave vector. In contrast, for the two-photon excitation to the Rydberg state, the Doppler shift acts as an effective random detuning in a periodically driven scheme, which accumulates from cycle to cycle and leads to phase errors in the collective dynamics. This effect is incorporated by introducing an additional error Hamiltonian $H_D = \sum_{i=1}^{\mathbb{N}} \delta_i^{D}\, |r_i\rangle\langle r_i|$, where $\delta_i^{D}$ denotes the Doppler-induced detuning experienced by the $i$th atom during the two-photon Rydberg excitation. To account for atomic motion, each atom is assigned a velocity randomly sampled from a one-dimensional Maxwell--Boltzmann distribution corresponding to a finite atomic temperature $T_a$. The Doppler shift is given by $\delta_i^{D} = k_{\mathrm{eff}} v_{\mathrm{rms}}$, where $k_{\mathrm{eff}} = |k_{780} - k_{480}|$ denotes the effective wave vector along the excitation ($z$) axis, determined by the difference in magnitude between the wave vectors of the counter-propagating fields.. The root-mean-square atomic velocity along this direction is $v_{\mathrm{rms}}=\sqrt{k_B T_a/m}$, where $T_a$ is the atomic temperature and $m$ is the atomic mass. We consider an atomic ensemble confined in a geometry where the motion transverse to the excitation axis is tightly constrained and can be neglected. Under this approximation, the dominant motional fluctuations arise from atomic velocities projected along the excitation ($z$) direction, which can be described as thermal velocity fluctuations superimposed on a slowly varying center-of-mass motion. Such one-dimensional treatments of atomic motion are commonly used in studies of Doppler dephasing and collective effects in cold atomic ensembles~\cite{PhysRevA.101.033602}. The Doppler-induced collective dephasing of the $|W_{\mathbb{N}}\rangle$ state is shown in Fig.~\hyperref[fig4]{4(e)} for $\mathbb{N}=4,5,6$ atoms at different temperatures. Each data point is obtained by averaging over 200 independent numerical realizations of atomic velocities drawn from the corresponding thermal distribution. For $T_a=5~\mu\mathrm{K}$, the final fidelity is reduced to approximately $96.1\%$, while increasing the temperature to $T_a=12~\mu\mathrm{K}$ further lowers the fidelity to approximately $91.6\%$, reflecting the cumulative nature of Doppler dephasing under periodic driving. In view of the Doppler-induced degradation discussed above, we employ an alternating driving protocol, shown in Fig.~\hyperref[fig4]{4(f)}, to reduce Doppler-induced effects. Reversing the propagation directions of the excitation beams in successive cycles flips the sign of the effective wave vector $k_{\mathrm{eff}}$, which causes the Doppler phase accumulated in one cycle to be partially compensated in the next. Consequently, long-term Doppler-induced dephasing is strongly suppressed, and the fidelity remains close to $99.0\%$ even at atomic temperatures up to $T_a = 12~\mu\mathrm{K}$, demonstrating the strong robustness of the proposed protocol against Doppler noise.
}

{
\subsection{Comparison of our scheme to previous schemes}

\begin{table}
\centering
\caption{Comparison of our scheme to previous schemes}
\label{tab1t}
\begin{tabular}{llll}
\toprule
Reference & Mechanism & System & Fidelity \\
\midrule
Our scheme & Floquet dynamics & atomic ensemble & $99\%$ (Theo.)  \\
Ref.~\cite{Jau2016} & Rydberg dressing & atomic array & $ 81\%$ (Exp.) \\
Ref.~\cite{PhysRevA.96.012328}  & Rydberg-antiblockade & atomic array, & $99\%$ (Theo.) \\
Ref.~\cite{PhysRevLett.125.073601} & induced dipole blockade  & regular chain & $96\%$ (Theo.)  \\
\bottomrule
\end{tabular}
\end{table}
In this section, we situate our work within the broader context of neutral-atom quantum logic by comparing our Floquet stroboscopic dynamics with several representative schemes that realize effective ground-state blockade through Rydberg-mediated mechanisms, including Rydberg dressing~\cite{Jau2016}, Rydberg antiblockade~\cite{PhysRevA.96.012328}, and light-induced dipole blockade~\cite{PhysRevLett.125.073601} as shown in Table~\ref{tab1t}.

First, the Rydberg dressing has been experimentally demonstrated as a powerful method for engineering effective interactions within the ground-state manifold~\cite{Jau2016}. In this approach, ground states are weakly and off-resonantly coupled to Rydberg states, producing tunable soft-core interactions. Nevertheless, a fundamental limitation of dressing schemes is the unavoidable admixture of the Rydberg character into the dressed states. As a result, the system is subject to continuous decoherence from the finite Rydberg lifetime, leading to an intrinsic trade-off between interaction strength and coherence time.

Subsequently, Ref.~\cite{PhysRevA.96.012328} proposed a scheme based on the Rydberg antiblockade mechanism~\cite{PhysRevA.96.012328}, in which the interaction-induced energy shift is tuned to compensate for the laser detuning. This effectively maps the blockade physics entirely onto the ground-state manifold, without requiring the population of real excited states, and endows the protocol with a certain degree of robustness. However, the principal limitation of this approach lies in its poor scalability rather than its sensitivity to noise. Because the van der Waals interaction scales as $1/R^6$, the antiblockade condition can only be satisfied in highly constrained geometries with well-controlled interatomic separations. Consequently, the scheme is restricted to small planar atomic arrays and has been demonstrated only for systems of up to three atoms, making it difficult to extend to larger atomic systems or ensembles.

Finally, Ref.~\cite{PhysRevLett.125.073601} investigated collective blockade effects mediated by dipole–dipole interactions. While this mechanism can induce photon or excitation blockade in dense atomic ensembles, it requires all atoms to be confined within a subwavelength volume ($R < \lambda/2\pi$). Such a stringent geometric constraint severely limits scalability and is incompatible with standard optical tweezer architectures, where individual addressability and readout typically require interatomic spacings on the order of micrometers.

In contrast, our Floquet-based protocol provides a robust and versatile route to realizing ground-state blockade with several key advantages. First, this ground-state blockade requires neither a predefined blockade radius nor strong vdW interactions, and it functions effectively without an additional cavity. Even in the regime of relatively weak Rydberg-Rydberg interactions, the effective ground-state blockade can still be established through Floquet engineering, substantially relaxing constraints on interatomic spacing and interaction strength. Second, the protocol is intrinsically robust against a wide range of experimental imperfections, including spontaneous emission, blackbody-radiation-induced decay, intermediate-state scattering, laser amplitude and phase noise, detuning errors, and Doppler-induced dephasing, as demonstrated by our numerical simulations. Finally, the blockade is effectively realized within the ground-state manifold at stroboscopic times, such that the system dynamics are well projected back onto the ground-state subspace at the end of each driving cycle. This feature effectively suppresses decoherence channels associated with finite Rydberg lifetimes and enables high-fidelity state preparation compatible with scalable, micrometer-spaced neutral-atom architectures.

}
\section{Single-photon source}\label{sec4}
\begin{figure}
	\centering
	\includegraphics[width=1\linewidth]{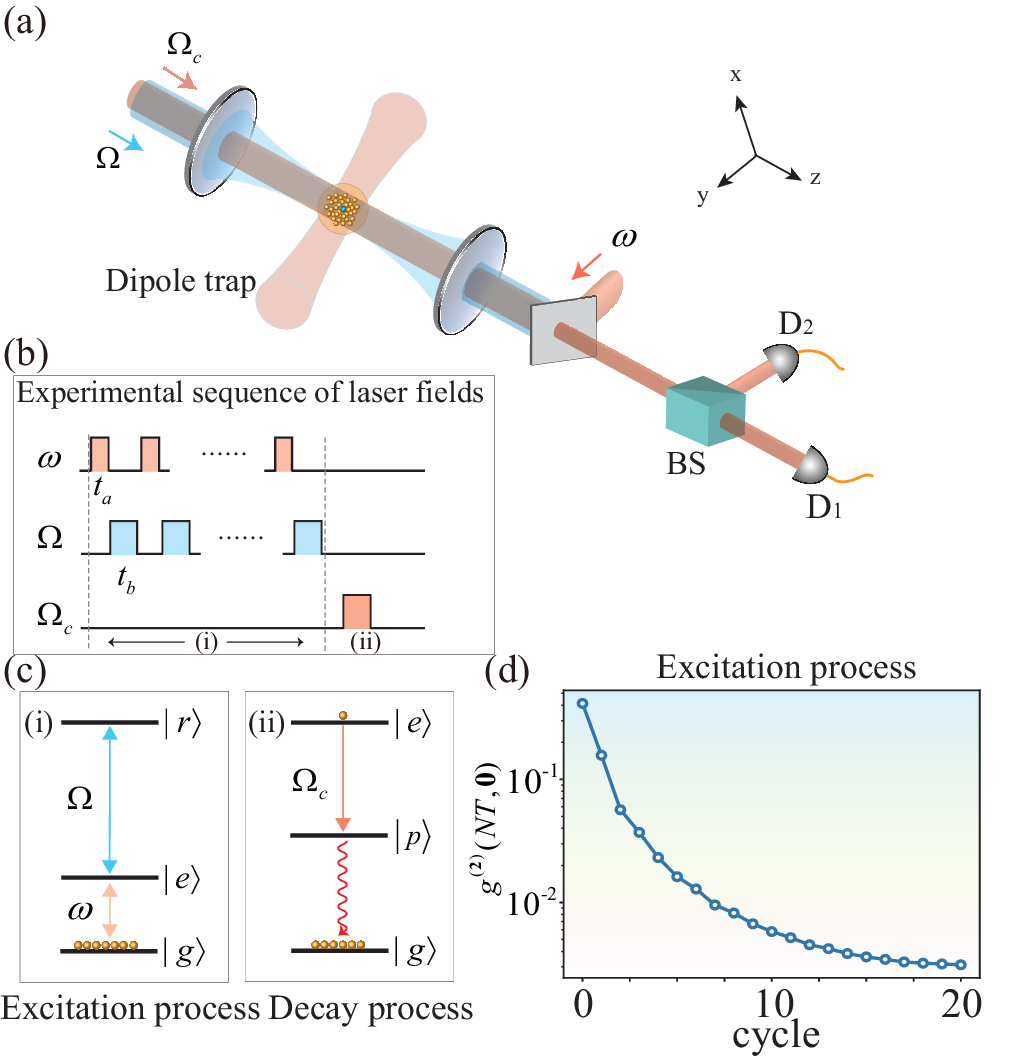}
	\caption{(a) An ensemble of cold $^{87}\text{Rb}$ atoms is confined in a dipole trap. The generated single photons are detected by detectors D1 and D2. (b) The evolution and corresponding pulse sequence of the system under periodic driving. (c) Level diagram. (i) Atoms are initially in the ground state $|g\rangle$ and prepared in the singly excited state $|e\rangle$ via optical pumping. (ii) The ground-state excitation is retrieved through the application of the read field $\Omega_c$, and subsequently measured at detectors D1 and D2. (d) The second-order correlation function of the radiated light. Other parameters are the same with Fig.~\ref{fig2}.}\label{fig5}
\end{figure}

Building upon the results presented above, we further explore their application to the preparation of a single-photon source, which mainly involves two key processes~\cite{PhysRevLett.121.123605,PhysRevA.109.013705,peyronel2012quantum,Ornelas-Huerta:20,li2016quantum}. The first is the excitation process, which prepares a single excitation state. This is followed by the emission process, where the single-excitation state is converted into a single photon via stimulated emission. This phenomenon provides a unique insight into the interplay between the atomic ensemble and the incident electromagnetic field. Notably, the formation of collective single-excitation states manifests itself as a distinctive signature in the light radiated, which is the most important step in the formation of single photons. The experimental schematic diagram is shown in Fig.~\hyperref[fig5]{5(a)}. The experimental sequence in Fig.~\hyperref[fig5]{5(b)} shows the pulse durations, while Fig.~\hyperref[fig5]{5(c)} illustrates the excitation and decay processes, with $\Omega_c$ facilitating the controlled readout pulse during the decay process. To quantitatively analyze the excitation process, we introduce the equal-time second-order correlation function~\cite{PhysRevLett.125.073601,PhysRevA.109.013710,PhysRevA.105.053715,PhysRevA.105.062408} 
\begin{align}
    g^{(2)}(t,\tau)\equiv \frac{\left\langle\hat{E}^{-}(t) \hat{E}^{-} (t+\tau)\hat{E}^{+}(t+\tau) \hat{E}^{+}(t)\right\rangle}{\left\langle\hat{E}^{-}(t) \hat{E}^{+}(t)\right\rangle^2},
\end{align}
which characterizes the system ability to emit two photons simultaneously at a time $t$, where $\hat{E}^{+} \sim \sum_{i=1}^\mathbb{N} e^{-i k \widehat{n} \cdot \mathbf{r}_i} \sigma_i^{-}$ refers to the radiated field in a direction $\widehat{n}$. $g^{(2)}(NT,0)< 1$ is a signal of the single-excitation nature of the state, the second-order correlation function briefly describes the statistical properties of photons. When $g^{(2)}(NT,0)> 1$, photon statistics describe the super-Poissonian distribution, and photons exhibit bunching effects. When $g^{(2)}(NT,0)< 1$, photons describe a sub-Poissonian distribution, and they exhibit antibunching effects. In Fig.~\hyperref[fig5]{5(d)}, we show numerical results for the second-order correlation function during the dynamical evolution of the system. It is evident that the value of the second-order correlation function $g^{(2)}(NT,0)$ decreases during the evolution process and reaches $10^{-3}$ after $20$ cycles, indicating an increased probability of the $W$ state. This is similar to the photon blockade mechanism, where the system emits only a single photon at a time, and its purity is determined by the occupancy probability of the $W$ state. After the collective excitation is prepared, the subsequent photon emission must be carefully controlled to enable single-photon output. Our approach offers a robust theoretical foundation for the preparation of single-photon sources.

\section{Implementation in a Rydberg Superatom}\label{sec5}

\begin{figure}
	\centering
	\includegraphics[width=1\linewidth]{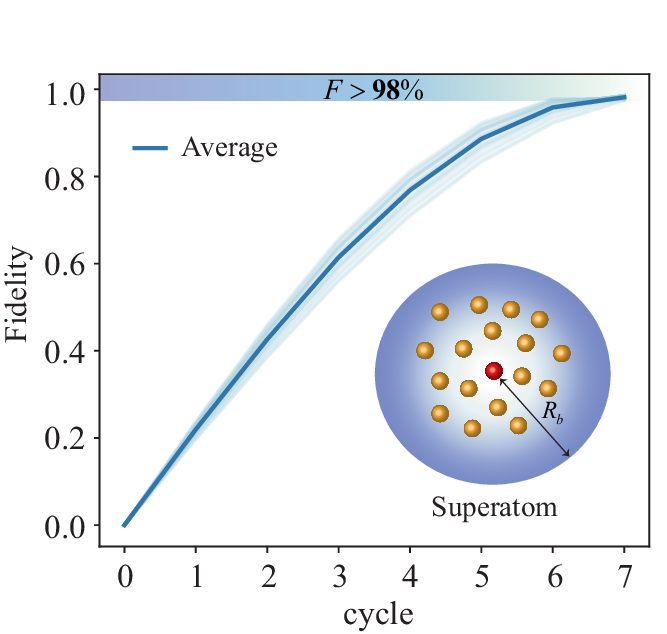}
	\caption{Evolution of the fidelity of the state $|W_\mathbb{N}\rangle$ under the analytical solution for a Rydberg superatom with $\mathbb{N} = 500$ atoms, together with the impact of $20\%$ atom number fluctuations. The parameters are the same with Fig.~\ref{fig2}.}\label{fig6}
\end{figure}

In principle, if the experimental techniques are sufficiently advanced to enable perfect Rydberg blockade and the impact of technical noise can be neglected, our protocol can be straightforwardly extended to systems with a large number of atoms~\cite{10.1063/5.0211071,Weber2015,PhysRevA.91.023813,PhysRevResearch.2.043339,https://doi.org/10.1002/qute.202200173}. In this idealized limit, the advantages of the scheme become even more pronounced, as the required number of driving cycles to prepare the target state decreases with increasing atomic number. To gain a deeper understanding of the experimental realization of ground-state blockade in a Rydberg superatom, we reinterpret the mechanism in terms of interaction-enhanced single-excitation transfer. Under periodic driving, the system dynamics allow only a single atom to undergo repeated oscillations between $|e\rangle$ and $|r\rangle$, while multi-atom participation is inhibited. Due to the strong Rydberg blockade that forbids the simultaneous excitation of multiple atoms in $|r\rangle$, this process effectively suppresses the occurrence of multiple $|e\rangle$ excitations. In other words, the periodic protocol acts as if it reinforces the mutual exclusion among $|e\rangle$ excitations, ensuring that only a single $|e\rangle$ remains. For completeness, we also provide the approximate analytical form of the system evolution under simplifying assumptions in Appendix~\ref{Appendix_B}.

As shown in Fig.~\hyperref[fig6]{6}, we numerically simulate the dynamics of the fidelity for the $|W_\mathbb{N}\rangle$ under different realizations of atom numbers $\mathbb{N}$ sampled from a predefined distribution. The fidelity is evaluated as $F(\mathbb{N}_0)=\langle\psi_{\mathrm{target}}(\mathbb{N}_0)|\rho_{\mathbb{N}_0}(t)|\psi_{\mathrm{target}}(\mathbb{N}_0) \rangle$, where $|\psi_{\mathrm{target}}(\mathbb{N}_0)\rangle=|W_{\mathbb{N}_0} \rangle$ corresponds to  the $W$ state for the $\mathbb{N}_0$ atoms. Here, $\mathbb{N}_0$ represents the actual atom number in a given experimental realization, which may deviate from the nominal number $\mathbb{N}$ due to atom loss.
This allows us to assess the robustness of the protocol against atom-number fluctuations. The pulse duration $t_a=\pi/(\sqrt{\mathbb{N}}N\omega)$ is designed for the reference atom number $\mathbb{N}=500$. The same control sequence, with this fixed pulse duration, is applied to all realizations with different actual atom numbers $\mathbb{N}_0$. Since the pulse duration is fixed, this deviation effectively manifests itself as a fluctuation in the effective evolution time rather than a change in the control protocol itself. The results demonstrate that even in the presence of $20\%$ fluctuations in atom number around $\mathbb{N}=500$, the system still converges to the symmetric collective state of the ground state with a high fidelity of $98\%$ after several driving cycles, highlighting the strong robustness of the scheme against atom number fluctuations.

\section{Corresponding Energy-Level Diagram and Transitions}\label{sec6}

\begin{figure}
	\centering
	\includegraphics[width=0.8\linewidth]{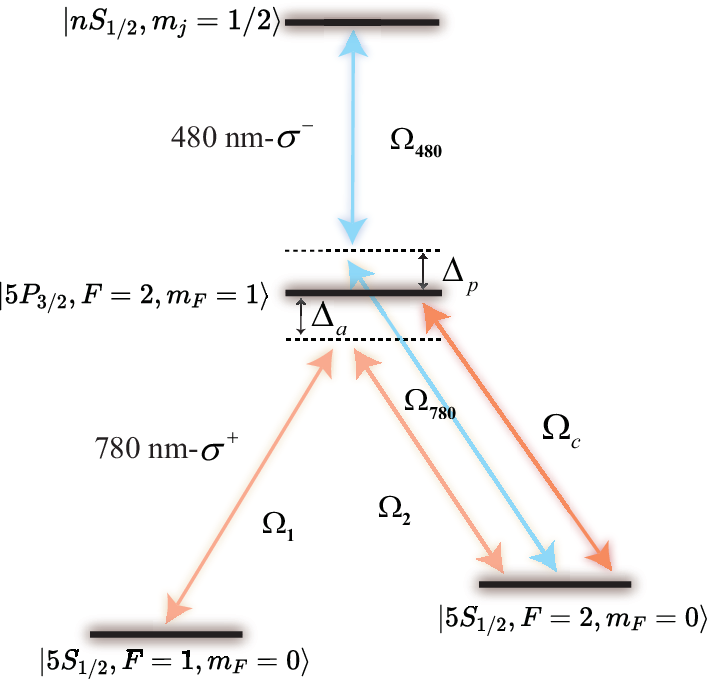}
	\caption{A schematic illustration of the $^{87}\text{Rb}$ atom showing the relevant energy level structure.}\label{fig7}
\end{figure}
In the experiment, we can find the atomic configuration required for our scheme in the $_{}^{87}$Rb atom, as shown in Fig.~\ref{fig7}. The ground states $|g\rangle$ and $|e\rangle$ correspond to the atomic levels $|5S_{1/2},F=1,m_F=0\rangle$ and $|5S_{1/2},F=2,m_F=0\rangle$, the Rydberg state is $|nS_{1/2},m_j=1/2\rangle $, $|5P_{3/2},F=2,m_F=1\rangle$ is an intermediate state that can be adiabatic eliminated under conditions $\Delta_a \gg \{ \Omega_1,\Omega_2\} $ and $\Delta_p \gg \{\Omega_{780},\Omega_{480} \}$, leading to a direct coupling between $|g\rangle $ and $|e\rangle$ with two-photon Rabi frequency $\omega$ and $|e\rangle$ and $|r\rangle$ with two-photon Rabi frequency $\Omega$. Ignoring the Stark shift, we can obtain the effective Hamiltonian of these two-photon processes as
\begin{align}
    H_a^\mathrm{eff}&=\frac{\Omega_1 \Omega_2}{4\Delta_a} |e\rangle \langle g|+\mathrm{H.c.},\notag\\
    H_b^\mathrm{eff}&=\frac{\Omega_{780} \Omega_{480}}{4\Delta_p} |r\rangle \langle e|+\mathrm{H.c.}   
\end{align}
The corresponding Rabi frequencies used in the text read $\omega=\Omega_1 \Omega_2/2\Delta_a$ and $\Omega=\Omega_{780} \Omega_{480}/2\Delta_p$. The Raman transition between the ground states $|g\rangle$ and $|e\rangle$ is driven by the use of two 780-nm laser beams with $\sigma^+$-polarized. The Rydberg excitation uses $\sigma^+$-polarized 780-nm and $\sigma^-$-polarized 480-nm beams tuned for excitation of the Rydberg state of $nS_{1/2}$. A wide range of experimentally accessible parameters can be used in our scheme. In particular, the Raman coupling between the ground states can be tuned over a broad interval, and values of the effective Rabi frequency $\omega/2\pi$ in the range of $0.5–3$ MHz are readily achievable with standard optical configurations~\cite{PhysRevLett.96.063001,Hattermann2017,PhysRevB.104.035201,PhysRevA.105.032618,PhysRevResearch.5.L012002,Smith2023}. Higher Rabi frequencies for the ground-to-Rydberg excitation are generally favorable, and current experimental techniques can provide sufficiently strong driving for our scheme~\cite{PhysRevLett.121.123603,evered2023high,Browaeys2020,w3x9-ll79,PhysRevResearch.4.013207}. For reference, Ref.~\cite{PhysRevLett.96.063001} demonstrated a two-photon Raman coupling between the states $|5S_{1/2},F=1,m_F=0\rangle$ and $|5S_{1/2},F=2,m_F=0\rangle$ with an effective Rabi frequency of $\omega/2\pi = 1.36$ MHz using an intermediate-state detuning of $\Delta_a/2\pi = -41$ GHz. In addition, Ref.~\cite{evered2023high} achieved a two-photon Rydberg excitation with intermediate-state detuning $\Delta_p/2\pi = 7.8$ GHz and a Rabi frequency of $\Omega/2\pi = 4.6$ MHz. In the single-photon generation scheme, we designate the state $|5P_{3/2},F=2,m_F=1\rangle$ as the controlled excited state, which is coupled to the excited state $|5S_{1/2},F=2,m_F=0\rangle$ via a $\sigma^+$-polarized 780-nm. Furthermore, the scheme does not depend on the strength of vdW interactions. From this perspective, the proposed mechanism is adaptable to a wide variety of parameters.

\section{Conclusion}\label{sec7}

We have proposed a Floquet-engineered scheme to enhance the effective interaction between ground-state neutral atoms in a Rydberg atomic ensemble. By periodically driving the system with a sequence of ground-state couplings followed by a $4\pi$ pulse on the ground–Rydberg transition, the dynamics are confined to the single-excitation manifold through stroboscopic evolution. As a result, the system evolves from the collective ground state $|G\rangle$ to the state $|W_\mathbb{N}\rangle$, while double excitations remain strongly suppressed. This behavior reflects an effective ground-state blockade induced by Floquet modulation. Unlike conventional schemes that rely on strong Rydberg-Rydberg interactions and are therefore limited to the blockade radius, our approach remains effective even outside the conventional Rydberg blockade radius. In addition, we evaluate the feasibility of the proposed scheme under realistic experimental conditions by systematically incorporating major sources of imperfections, including Rydberg-state spontaneous emission and blackbody-radiation-induced decay, laser phase noise, Rabi amplitude fluctuations, detuning errors, as well as Doppler-induced dephasing arising from finite atomic temperature. Our results demonstrate that the scheme exhibits strong robustness against these noise sources and remains resilient to thermal motion and ensemble inhomogeneity, indicating that it is well suited for implementation in current neutral-atom platforms. We note that additional noise sources specific to particular experimental implementations may also contribute to decoherence but are beyond the scope of the present theoretical model. Building on this mechanism, we further examine its application to single-photon sources. The simulated equal-time second-order correlation function indicates that the scheme can theoretically achieve a high-efficiency single-photon source. Finally, we explore the implementation of our scheme in a Rydberg superatom, demonstrating that shortening the Floquet driving period enables practical realization within a single Rydberg superatom. In summary, our scheme realizes robust quantum state preparation by effectively enhancing ground-state interactions in neutral atoms. This mechanism can be directly applied to deterministic single-photon generation, providing a versatile and high-fidelity platform for quantum technologies.

\begin{acknowledgments}
This work is supported by the National Natural Science Foundation of China (Grants No.~12174048 and 62375047), the Italian PNRR MUR (No.~PE0000023-NQSTI), I-PHOQS (Photonics and Quantum Sciences, PdGP/GePro 2024-2026), and the Fund for International Activities of the University of Brescia.
\end{acknowledgments}

\section*{Data Availability}
The data that support the findings of this study are openly available in \cite{wei2025enhancing}.

\appendix

\renewcommand{\theequation}{A\arabic{equation}} 
\setcounter{equation}{0}

\section{The evolution of quantum states under unitary kicks}\label{Appendix_A}
The time evolution of the system is generated by a periodic sequence of unitary operators, which alternate between free evolution and kicks unitary transformation~\cite{SHAO20111099,morton2006bang}. The dynamic evolution operator of the system is as follows:
\begin{align}\label{eqa1}
U_N (t_a+t_b)=[ U_{b}(t_b) U_{a} (t_a)]^N,
\end{align}
$U_a(t_a)$ represents the free evolution operator of the Hamiltonian $H_a$, and $U_{b} (t_b)$ denotes the kicks unitary transformation of the Hamiltonian $H_b$, with the duration of one cycle given by $T=t_a+t_b$. Consequently, we can express this evolution operator as follows:
\begin{align}\label{eqa2}
U_N (T)& = U_{b} (t_b)^N U_{b} (t_b)^{\dagger N} U_N (t_a+t_b)\notag \\
& =U_{b} (t_b)^N V_N{(t_a)},
\end{align}
where $V_N{(0)}=I$, and

\begin{align}\label{eqa3}
i \frac{d}{d t_a} V_N (t_a) & =U_{b}^{\dagger N} \sum_{k=0}^{N-1}\left(U_{ba}\right)^k\left(U_{b} i \frac{d U_{a}}{d t}\right)\left(U_{ba}\right)^{N-k-1}\notag \\
& =U_{b}^{\dagger N} \frac{1}{N} \sum_{k=0}^{N-1}\left(U_{ba}\right)^k U_{b} H_a U_{b}^{\dagger}\left(U_{ba}\right)^{\dagger k}\left(U_{ba}\right)^N\notag \\
& =H_N(t_a) V_N(t_a)
\end{align} 
with 
\begin{align}\label{eqa4}
H_N(t_a)=\frac{1}{N} \sum_{k=0}^{N-1} U_{b}^{\dagger N}\left(U_{ba}\right)^k U_{b} H_a U_{b}\left(U_{ba}\right)^{\dagger k} U_{b}^N,
\end{align}
where $U_{ba}=U_b U_a$. In the large limit $N$, $H_N(t_a) \sim  1/N \sum_{k=0}^{N-1} U_{a}^{\dagger k} H_a U_{a}^{ k} $, the limiting evolution operator is
\begin{align}\label{eqa5}
U_{a}\left(t_a\right)=\lim _{N \rightarrow \infty} \mathcal{V}_N\left(t_a\right) \sim \exp \left(-i H_z t_a\right),
\end{align}
where 
\begin{align}\label{eqa6}
H_z=\lim_{N \to \infty}H_N(t_a)= \sum_{n} P_n H_a P_n,
\end{align}
is called Zeno Hamiltonian. $P_n$ represents the spectral projection of $U_b$ corresponding to the eigenvalue $\lambda_n$, with $U_b (t_b)= \sum_{n}\exp(-i \lambda _n t_b) P_n$. By combining Eq.~(\ref{eqa2}) and Eq.~(\ref{eqa5}), the whole system is governed by the effective evolution operator:
\begin{align}\label{eqa7}
\lim _{N \rightarrow \infty}U_N(T) &\equiv \mathcal{U}_N (T)\notag \\
& \sim U_{b} (t_b )^N e^{-i H_z N t_a}\notag  \\
&=\exp \left[-i \sum_n (\lambda_n P_n \frac{t_b}{T} + P_n H_a P_n \frac{t_a}{T}) N T\right].
\end{align}
Therefore, the effective Hamiltonian is described
\begin{align}\label{eqa8}
H_e= \sum_n \lambda_n P_n \frac{t_b}{T} + P_n H_a P_n \frac{t_a}{T}.
\end{align}

\renewcommand{\theequation}{B\arabic{equation}} 
\setcounter{equation}{0}
\section{The Zeno Hamiltonian of two atoms and dynamics of multi-atomic symmetric state}\label{Appendix_B}
For the effective Zeno Hamiltonian of two atoms, we also consider the following five basis vectors in the presence of the Rydberg blockade effect: \{ $|gg\rangle$, $|W_2\rangle=(|ge\rangle+|eg\rangle)/\sqrt{2}$, $|ee\rangle$, $|T_2\rangle=(|gr\rangle+|rg\rangle)/\sqrt{2}$, $|D_2\rangle=(|re\rangle+|er\rangle)/\sqrt{2}$ \}. Thus, the Hamiltonian $H^{(1)}_{a}$ and $H^{(1)}_{b}$ in the basis vectors are as follows:
\begin{align}
H^{(1)}_{a} =
\frac{1}{2}\begin{bmatrix}
0 & \sqrt{2} \omega & 0 & 0 & 0 \\
\sqrt{2} \omega & 0 & \sqrt{2} \omega & 0 & 0 \\    
0 & \sqrt{2} \omega & 0 & 0 & 0 \\
0 & 0 & 0 & 0 & \omega \\
0 & 0 & 0 & \omega & 0
\end{bmatrix}
\end{align}
and
\begin{align}
H^{(1)}_{b} &= 
\frac{1}{2}\begin{bmatrix}
0 & 0 & 0 & 0 & 0 \\
0 & 0 & 0 & \Omega & 0 \\    
0 & 0 & 0 & 0 & \sqrt{2} \Omega \\
0 & \Omega & 0 & 0 & 0 \\
0 & 0 & \sqrt{2} \Omega & 0 & 0
\end{bmatrix}.
\end{align}
Therefore, we can calculate the eigenvalues of the evolution operator $U_b(t_b)=e^{-iH^{(1)}_{b} t_b}$ corresponding to the Hamiltonian $H^{(1)}_{b}$: $e^{2i \sqrt{2}\pi}$, $e^{-2i \sqrt{2}\pi}$, 1, 1, and 1. The corresponding eigenstate is
\begin{gather}
|\phi_1\rangle=\frac{\sqrt{2}}{2}(|ee\rangle+|T_2\rangle),\notag\\
|\phi_2\rangle=\frac{\sqrt{2}}{2}(|ee\rangle-|T_2\rangle),\notag\\
|\phi_3\rangle=|gg\rangle,\notag\\
|\phi_4\rangle=|W_2\rangle,\notag\\
|\phi_5\rangle=|D_2\rangle,
\end{gather}
then the corresponding projection operators $P_1=|\phi_1\rangle \langle \phi_1|$, $P_2=|\phi_2\rangle \langle \phi_2|$, and $P_3=|\phi_3\rangle \langle \phi_3|+|\phi_4\rangle \langle \phi_4|+|\phi_5\rangle \langle \phi_5|$ result in the Zeno Hamiltonian as follows:
\begin{align}
H_z &=P_1 H^{(1)}_{a} P_1+P_2 H^{(1)}_{a} P_2+P_3 H^{(1)}_{a} P_3 \notag\\
 &=
\frac{1}{2}\begin{bmatrix}
0 & \sqrt{2} \omega & 0 & 0 & 0 \\
\sqrt{2} \omega & 0 & 0 & 0 & 0 \\    
0 & 0 & 0 & 0 & 0 \\
0 & 0 & 0 & 0 & 0 \\
0 & 0 & 0 & 0 & 0
\end{bmatrix}\notag\\
&=\frac{\sqrt{2} \omega}{2} ( |gg\rangle \langle W_2|+|W_2\rangle \langle gg| ).
\end{align}
Therefore, the effective Hamiltonian of this zeno subspace is $H_z^{\text{eff}}=H_z t_a/(t_a+t_b)$, and the effective kick Hamiltonian is $H_k^{\text{eff}}=H^{(1)}_{b} t_b/(t_a+t_b)$.

For the Rydberg superatom case, we derive the first-step evolution operator by considering the small $\sqrt{\mathbb{N}}\omega t$ approximation. The analysis begins by examining the single-atom evolution operator under the first resonant pulse with Rabi frequency $\omega$, described by the unitary operator $|\phi(t)\rangle=\cos(\omega t/2) |g\rangle-i \sin(\omega t/2) |e\rangle$. In the absence of interatomic interactions, the time evolution of a multi-atom system can be approximated by the tensor product of individual single-atom evolution operators. In the short-time limit $(\sqrt{\mathbb{N}}\omega t \ll 1)$, the dynamics of the system, initially in the ground state $|G\rangle = |g_1 g_2 \dots g_\mathbb{N}\rangle$, can be expressed through the second-order perturbation expansion as the time evolution.
\begin{align}
|\Psi(t)\rangle \propto & |G\rangle - i\frac{\omega t}{2} \sqrt{\mathbb{N}}|W_\mathbb{N}\rangle \notag \\ 
&- (\frac{\omega t}{2})^2 \left[ \frac{\mathbb{N}}{2}|G\rangle + \sqrt{\frac{\mathbb{N}(\mathbb{N}-1)}{2}}|P_\mathbb{N}\rangle \right] + \mathcal{O}((\omega t)^3),
\end{align}
where $|P_\mathbb{N}\rangle=\tfrac{1}{\sqrt{C_{\mathbb{N}}^{2}}}\sum_{j<k}|g\ldots e_{j}\ldots e_{k}\ldots g\rangle$. This evolution operator reveals the collective quantum behavior of a multiatom system: Under weak driving ($\sqrt{\mathbb{N}}\omega t \ll 1$), the system exhibits a collectively enhanced transition from the ground state $|G\rangle$ to the state $|W_\mathbb{N}\rangle$ with an effective Rabi frequency $\sqrt{\mathbb{N}} \omega$. In the next order, two-photon processes proportional to $t^2$ populate the double-excitation state $|P_\mathbb{N}\rangle$ and contribute a second-order correction back to the ground state, marking the onset of higher-order collective dynamics. The entire expression exhibits characteristic many-body effects, such as collective enhancement and hierarchical transitions, reflecting a perturbative expansion of typical Dicke-model dynamics. Then, taking into account the Rydberg blockade effect, which restricts the system to at most one excitation in the Rydberg state $|r\rangle$, we construct the corresponding complete basis as follows:
\begin{align}
    |G\rangle &=|g_1 g_2...g_\mathbb{N} \rangle \quad \notag \\
    |W_\mathbb{N}\rangle &= \frac{1}{\sqrt{\mathbb{N}}}\sum_{i=1}^\mathbb{N} |g_1\cdots e_i\cdots g_\mathbb{N}\rangle \quad \notag \\
    |P_\mathbb{N}\rangle &= \frac{1}{\sqrt{C_{\mathbb{N}}^{2}}}\sum_{i<j} |g_1\cdots e_i\cdots e_j\cdots g_\mathbb{N}\rangle \quad \notag \\
    |T_\mathbb{N}\rangle &= \frac{1}{\sqrt{\mathbb{N}}}\sum_{i=1}^\mathbb{N} |g_1\cdots r_i\cdots g_\mathbb{N}\rangle \quad \notag \\
    |D_\mathbb{N}\rangle &= \frac{1}{\sqrt{\mathbb{N}(\mathbb{N}-1)}}\sum_{i\ne j} |g_1\cdots e_i\cdots r_j\cdots g_\mathbb{N}\rangle  \quad 
\end{align}
Therefore, the two-step Hamiltonian under Rydberg blockade condition is as follows:
\begin{widetext}
\begin{align}
H^S_a &= \frac{1}{2}
\begin{pmatrix}
0 & \sqrt{\mathbb{N}}\omega     & 0 & 0 & 0 \\
\sqrt{\mathbb{N}}\omega     & 0 & \sqrt{2(\mathbb{N}-1)}\omega & 0 & 0 \\
0 & \sqrt{2(\mathbb{N}-1)}\omega & 0 & 0 & 0 \\
0 & 0 & 0 & 0 & \sqrt{(\mathbb{N}-1)}\omega \\
0 & 0 & 0 & \sqrt{(\mathbb{N}-1)}\omega & 0
\end{pmatrix},
&
H^S_b &= \frac{1}{2}
\begin{pmatrix}
0 & 0 &0 &0 &0 \\
0 & 0 &0 &\Omega &0 \\
0 & 0 &0 &0 &\sqrt{2}\Omega \\
0 &\Omega &0 &0 &0 \\
0 & 0 &\sqrt{2}\Omega &0 &0
\end{pmatrix}.
\end{align}
\end{widetext}

To evaluate the robustness of the protocol against atom-number fluctuations, we define fidelity as
\begin{align}
F(\mathbb{N}_0) =\langle\psi_{\mathrm{target}}(\mathbb{N}_0)|\rho_{\mathbb{N}_0}(t)|\psi_{\mathrm{target}}(\mathbb{N}_0) \rangle.
\end{align}
The target state $\big|\psi_{\mathrm{target}}(\mathbb{N}_0)\big\rangle$ is chosen as the symmetric collective state
\begin{align}
\big| \psi_{\mathrm{target}}(\mathbb{N}_0) \big\rangle 
= \big| W_{\mathbb{N}_0} \big\rangle 
= \frac{1}{\sqrt{\mathbb{N}_0}} \sum_{i=1}^{\mathbb{N}_0} 
\big| g_1 g_2 \cdots e_i \cdots g_{\mathbb{N}_0} \big\rangle.
\end{align}
In the simulation, the reference atom number is fixed at $\mathbb{N}=500$, which is used to determine the pulse parameters and the evolution time $t_a=\pi/(\sqrt{\mathbb{N}} N \omega)$. For each realization, the actual number of atoms $\mathbb{N}_0$ is randomly sampled from the range $\mathbb{N}_0 \in [400,600]$, and the fidelity is independently evaluated in the corresponding Hilbert space of $\mathbb{N}_0$ atoms. This procedure captures the impact of atom-number fluctuations while keeping the control sequence unchanged.

\renewcommand{\theequation}{C\arabic{equation}} 
\setcounter{equation}{0}
\section{Visualization of Continuous-Time Evolution}\label{Appendix_C}
\begin{figure}
	\centering
	\includegraphics[width=1\linewidth]{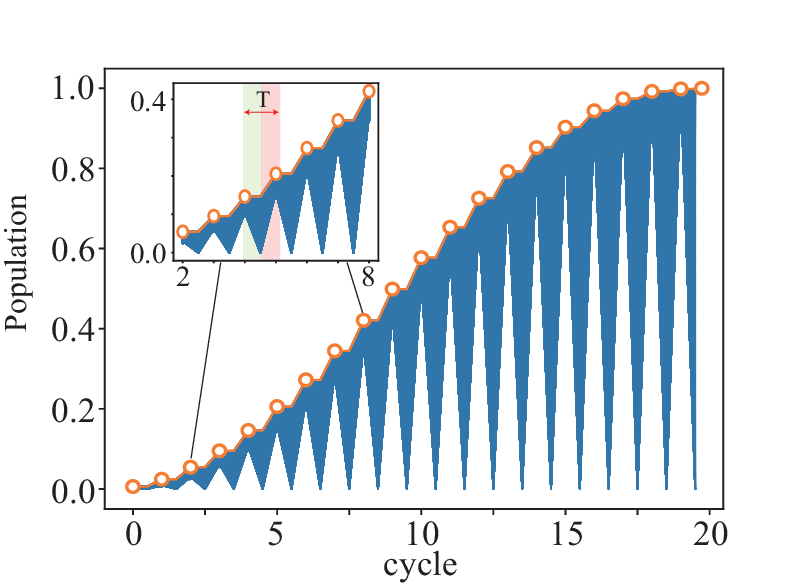}
	\caption{Continuous-time evolution of the state $|W_2\rangle$.}\label{fig8}
\end{figure}

In this paper, we present the stroboscopic dynamics after each driving cycle ends. To provide a comprehensive perspective on the dynamic continuous evolution of the system, we present the continuous-time evolution population diagram of the state $|W_2\rangle$ in Fig.~\ref{fig8}. As illustrated in this figure, we can intuitively observe the evolutionary behavior of the system at different time scales and reveal the quantum state changes throughout the entire evolution process.
This is because, during the second step of the pulse sequence, the atom transitions from the excited state $|e\rangle$ to the Rydberg state $|r\rangle$, forming a dressed state. In the single photon generation scheme, since excitation is selectively applied to the state $|e\rangle$, it is not influenced by the presence of the Rydberg state and can still lead to the emission of single photons. This visualization helps to understand evolutionary characteristics across various time scales and serves as a valuable reference for analyzing the underlying physical mechanisms.

\renewcommand{\theequation}{D\arabic{equation}} 
\setcounter{equation}{0}
\section{Fidelity Analysis for Three- and Four- Atom Systems}\label{Appendix_D}

\begin{figure}
	\centering
	\includegraphics[width=1\linewidth]{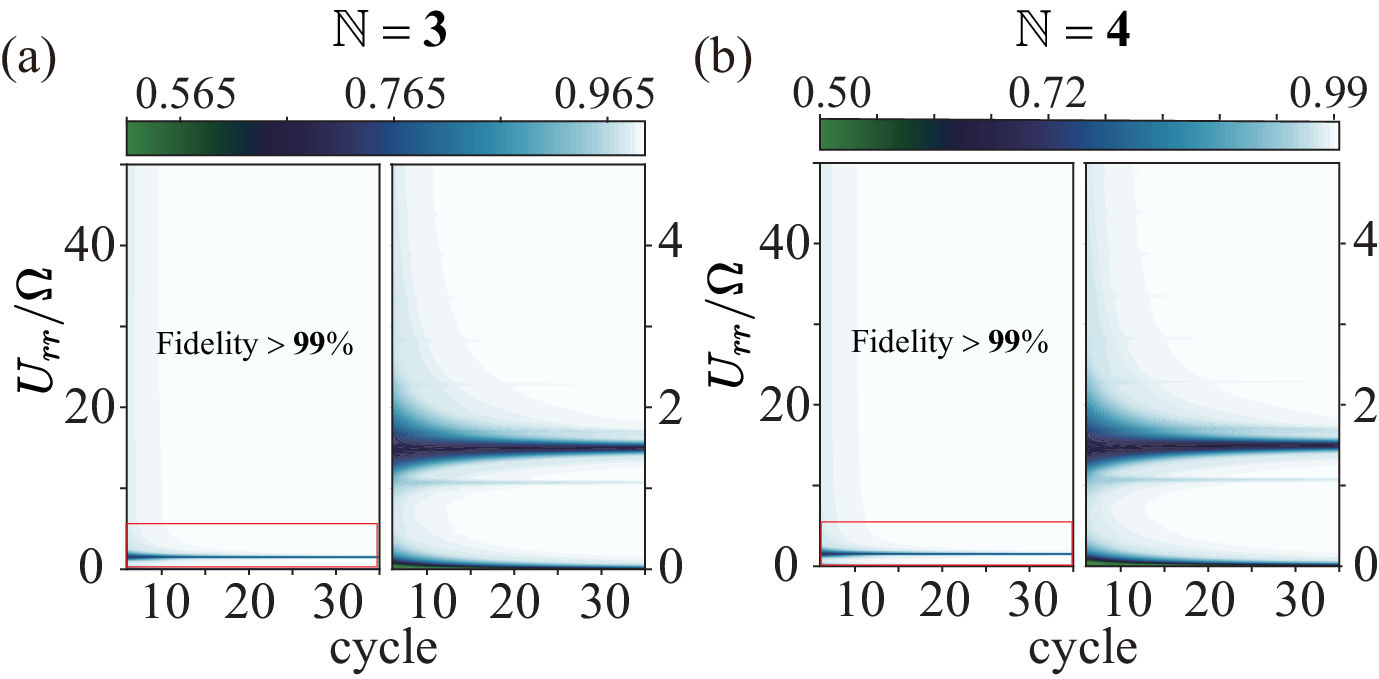}
	\caption{Fidelity of the state $|W_\mathbb{N}\rangle$ versus $U_{rr}/\Omega$ and cycle number $N$ for systems of (a) three and (b) four atoms, showing high overall robustness ($F>99\%$) but revealing several sharp, $N$-independent resonant leakage channels, as shown in the broad scan (i) and detailed view (ii).}\label{fig9}
\end{figure}

To verify the scalability and robustness of the proposed scheme, we extend the analysis to systems of three and four atoms. Figures.~\hyperref[fig9]{9(a)} and ~\hyperref[fig9]{9(b)} present the fidelity of the target state $|W_\mathbb{N}\rangle$ as a function of the Rydberg interaction strength $U_{rr}/\Omega$ and the cycle number $N$ for the three- and four-atom cases, respectively. The general behavior remains consistent with the two-atom case: high fidelities exceeding $99\%$ are achieved across most of the parameter space, demonstrating strong robustness against variations in the interaction strength and system size. Similar $N$-independent vertical leakage channels are observed near the same resonance positions, indicating that these resonant features are intrinsic to the dynamical mechanism rather than specific to a particular atom number. No additional degradation or broadening of the leakage structures is found, confirming that the scheme maintains its performance for larger atomic ensembles.

Importantly, because of the continuous variation in interatomic distances caused by thermal motion, the instantaneous Rydberg interaction strengths experienced by different atoms fluctuate from cycle to cycle, preventing all atoms from simultaneously occupying these resonant leakage conditions. As a result, the overall state preparation remains highly robust even under imperfect blockade conditions, reinforcing the scalability and feasibility of the scheme for multi-atom systems.

\bibliography{manuscript.bbl}

\end{document}